\documentclass[12pt,nofootinbib,preprintnumbers]{revtex4-1}
\usepackage{graphicx}
\usepackage{amsmath,amsthm,amssymb,multirow}
\usepackage{xcolor}
\usepackage{caption}
\usepackage{subcaption}
\usepackage[utf8]{inputenc} 
\usepackage{amssymb}
\usepackage{enumerate}
\usepackage{epstopdf}
\usepackage{multirow}
\usepackage{hyperref}
\usepackage{appendix}
\usepackage{url}
\usepackage{caption}
\usepackage{subcaption}
\usepackage{wasysym}
\usepackage{tabulary}
\usepackage{slashed}

\makeatletter 
\gdef\@ptsize{2}
\let\@currsize\normalsize 
\makeatother 

\usepackage{setspace}
\usepackage{tabularx}

\newcommand{\alignStart}{ \begin{equation} \begin{aligned} }
\newcommand{\alignEnd}{ \end{aligned} \end{equation} }
\newcommand{\gatherStart}{ \begin{equation} \begin{gathered} }
\newcommand{\gatherEnd}{ \end{gathered} \end{equation} }

\newcommand{\mpl}{M_\text{pl}}
\newcommand{\geff}{g_\text{eff}}
\newcommand{\eg}{g}
\newcommand{\eeff}{g_\text{eff}}
\newcommand{\feff}{f_\text{eff}}
\newcommand{\nefolds}{{\cal N}_\text{e-folds}}

\begin{document}

\title{The Weak Gravity Conjecture and Effective Field Theory}
  
\author{Prashant Saraswat}
\affiliation{Maryland Center for Fundamental Physics, Department of Physics, University of Maryland, College Park, MD 20742}
\affiliation{Department of Physics and Astronomy, Johns Hopkins
  University, Baltimore, MD 21218, USA}

\preprint{UMD-PP-016-007}
\begin{abstract}
The Weak Gravity Conjecture (WGC) is a proposed constraint on theories with gauge fields and gravity, requiring the existence of light charged particles and/or imposing an upper bound on the field theory cutoff $\Lambda$. If taken as a consistency requirement for effective field theories (EFTs), it rules out possibilities for model-building including some models of inflation. I demonstrate simple models which satisfy all forms of the WGC, but which through Higgsing of the original gauge fields produce low-energy EFTs with gauge forces that badly violate the WGC. These models illustrate specific loopholes in arguments that motivate the WGC from a bottom-up perspective; for example the arguments based on magnetic monopoles are evaded when the magnetic confinement that occurs in a Higgs phase is accounted for. This indicates that the WGC should not be taken as a veto on EFTs, even if it turns out to be a robust property of UV quantum gravity theories. However, if the latter is true then parametric violation of the WGC at low energy comes at the cost of non-minimal field content in the UV. I propose that only a very weak constraint is applicable to EFTs, $\Lambda \lesssim \left(\log \frac{1}{\eg}\right)^{-1/2} \mpl$ where $g$ is the gauge coupling, motivated by entropy bounds. Remarkably, EFTs produced by Higgsing a theory that satisfies the WGC can saturate but not violate this bound.
\end{abstract} 
\maketitle

\section{Introduction \& Synopsis}
\label{sec:intro}

While local quantum field theory has been a remarkably successful framework for fundamental physics, we remain largely ignorant of the nature of physics far in the ultraviolet (UV) where quantum gravity effects become strong. However, we may still make progress in explaining phenomena at lower energies using effective field theories (EFTs), where one remains agnostic about the detailed dynamics above some cutoff scale (though often assuming certain symmetries of the UV physics). 

The unspoken assumption in this approach is that the EFT under consideration could emerge at low energies from a complete theory of quantum gravity (QG). This assumption is challenged by the ``swampland" hypothesis~\cite{Vafa:2005ui, Ooguri:2006in}, which suggests that some EFTs cannot be consistently completed into a UV quantum gravity theory. Broadly, there are two approaches to addressing this issue: from the ``top-down," one may explore candidate QG theories such as string theory and attempt to make generic statements about the set of low-energy EFTs that can emerge from it (distinguishing the ``landscape" from the ``swampland"). This cannot be exhaustive however, so to truly rule out an EFT one must take a ``bottom-up'' approach of directly arguing that it is somehow inconsistent with QG (see e.g.~\cite{Adams:2006sv} for an example of a subtle breakdown of a na\"{i}vely valid EFT).

The Weak Gravity Conjecture (WGC)~\cite{WGC} is an example of a proposed constraint on the possible gauge EFTs that could arise from quantum gravity. In fact there are many different specific forms of this conjecture that have been proposed~\cite{WGC, Cheung:2014vva, Heidenreich:2015wga, Heidenreich:2015nta, Hebecker:2015zss, Heidenreich:2016aqi}. These generally either mandate the existence of charged particles with mass less than their charge in Planck units ($m < q \mpl$), or bound the cutoff scale $\Lambda$ of a gauge EFT as $\Lambda \lesssim \eg \mpl$ where $\eg$ is the gauge coupling. Similarly there have been various different arguments in support of these conjectures. From the top-down, numerous examples have been presented of the WGC criterion being enforced in examples from string theory~\cite{WGC, Hebecker:2015zss} or models of emergent gauge fields~\cite{Harlow:2015lma}. Bottom-up arguments typically make use of universal, low-energy features of theories with gauge fields and gravity, namely charged black holes (BHs) (though see also the results of~\cite{Cheung:2014ega} based on requirements of unitarity, analyticity and causality of low-energy theories.)

The constraints from the Weak Gravity Conjecture are particularly relevant for models of cosmic inflation. This is best illustrated by the example of the ``extranatural inflation" model~\cite{extraNatural}, in which the inflaton $\phi$ arises from the Wilson line of a $U(1)$ gauge field around an extra dimension, a $S^1$ of radius $R$. Gauge invariance and compactness of the $U(1)$ group imply that the potential can include only terms of the form $\cos \frac{n \phi}{f}$ with $f = \left( 2\pi R \eg\right)^{-1}$ and $n \in \mathbb{Z}$. Inflation then lasts for $\nefolds \sim f/\mpl$ $e$-foldings, so if the gauge coupling $\eg$ can be taken arbitrarily small then this model can produce an arbitrarily long period of inflaton. However, the WGC would indicate that the cutoff of this 5D gauge theory must be appear at a scale $\Lambda < \eg \mpl$. Requiring $\Lambda < 1/R$ to maintain some regime of 5D EFT control then bounds the inflaton field range $f \lesssim \mpl$, preventing inflation. This behavior is similar to that found for string theory axions-- field ranges are typically bounded to be sub-Planckian in controlled settings~\cite{Banks:2003sx}. The viability of trans-Planckian field excursions is also relevant to models involving new axion-like particles, such as in the recent framework of cosmological relaxation of the weak scale~\cite{Graham:2015cka}.

Thus there are important consequences for model-building if the Weak Gravity Conjecture is taken as a ``veto" on EFTs. Of course, the WGC is a conjecture and not a theorem, meaning that there is no proof of any inconsistency when it is violated. However, between the wealth of top-down examples in which it is satisfied and the various bottom-up arguments put forward to motivate it more universally, one might plausibly imagine that it is indeed satisfied in any EFT emerging from quantum gravity. 

In this paper however, I argue instead that while the WGC may correctly reflect a constraint on how gauge theories UV complete into quantum gravity theories, this does not imply that low-energy EFTs satisfy such constraints. 
Specifically, I will consider a toy model which satisfies the various forms of the WGC, and show that spontaneous symmetry breaking can then lead to a low energy EFT with a gauge force that strongly violates the WGC. Therefore, from an EFT perspective we cannot take the WGC as a criterion to rule out models, as a model that violates the WGC could be UV-completed into one that satisfies it, without affecting the low-energy theory. One can then ask about the status of the various bottom-up arguments that have been made to motivate the WGC for general gauge EFTs. The constructions I consider shed light on specific loopholes to these arguments that nullify their ability to constrain EFTs.

Within the models I consider, parametric violation of the WGC comes at the price of specific non-minimal particle content, with increasingly severe violation requiring more contrived models. This could suggest that in a possible landscape of EFTs the WGC is ``generically'' satisfied, and that moderate violations of the WGC are more common than extreme violations. 

While the standard parametrics of the WGC bound can be badly violated in these models, they do not in fact allow arbitrarily small gauge couplings if one accounts for the tendency of large numbers of species to lower the cutoff (for fixed Planck scale). Therefore one cannot realize the limit of an exact global symmetry (zero gauge coupling), which is generally deemed to be in conflict with black hole quantum mechanics and/or entropy bounds. Furthermore, in section~\ref{sec:ultimate} I argue that entropy bounds (e.g. the Bekenstein bound~\cite{Bekenstein:1980jp} or holographic entropy bounds~\cite{Fischler:1998st,Bousso:1999xy}) give a direct, bottom-up constraint on gauge EFTs of the form $\Lambda \lesssim \left(\log \frac{1}{\eg}\right)^{-1/2} \mpl$. Interestingly, the Higgsing models I discuss can saturate, but not violate, this bound if the WGC is satisfied in the UV theory. This bound is therefore a better candidate for an inescapable constraint on gauge EFTs than the conventional WGC $\Lambda \lesssim \eg \mpl$.

This paper is organized as follows. In section~\ref{sec:arguments}, I review the common formulations of the Weak Gravity Conjecture and the arguments involving black holes that have been presented to motivate them. In section~\ref{sec:higgs}, I present a model in which Higgsing produces a low-energy EFT which violates the WGC, and discuss how this EFT in fact manages to avoid the potential problems with black hole physics suggested by the previous arguments. In section~\ref{sec:ultimate}, I propose the weaker bound on gauge EFTs discussed above, motivated both by the limits of the Higgsed models I discuss and by considerations of black hole entropy. Finally, I conclude in section~\ref{sec:conc} with some discussion of the role of the WGC in guiding model-building in light of these results.

\section{Bottom-up arguments for a Weak Gravity Conjecture}
\label{sec:arguments}

The term ``Weak Gravity Conjecture'' encompasses a number of different proposed constraints~\cite{WGC, Cheung:2014vva, Heidenreich:2015wga, Heidenreich:2015nta, Hebecker:2015zss, Heidenreich:2016aqi} on gauge theories coupled to gravity, generally involving the parametric scale $\eg \mpl$ (for a single $U(1)$ field with coupling $\eg$) but differing in their precise physical implications. As a broad categorization we can distinguish between ``electric'' forms of the WGC, which make requirements on the spectrum of charged particles in the theory, and the ``magnetic'' WGC, which requires that effective field theory break down at or below the scale $\Lambda \sim \eg \mpl$ (motivated by arguments involving magnetic monopoles). I now review these statements of the WGC and critique the bottom-up arguments motivating them, which will offer insight into how they are evaded by the model of section~\ref{sec:higgs}.

\subsection{``Electric'' form}
\label{sec:electric}

The most minimal electric form of the WGC simply requires that any $U(1)$ gauge theory contain a charged particle of mass $m$ and charge $q$ such that $m < q$, in ``GR units" where Newton's constant and electric charge are normalized such that extremal black holes have $M = Q$.~\footnote{Though I will use these units throughout, most of the expressions I will write besides this extremality condition will be only \emph{parametric} relations, without $O(1)$ or even $O(8 \pi)$ factors included. Also, I will often write $\mpl$ explicitly, even though $\mpl \sim O(1)$ in these units.} This is a necessary condition for extremal black holes to be kinematically able to decay completely into fundamental charged particles~\cite{WGC}. It has been argued that exact stability for extremal black holes is pathological in the same way as conjectured stable black hole ``remnants" invoked to address the black hole information paradox~\cite{Aharonov:1987tp, Susskind:1995da}, and that this motivates the conjecture.

There are two immediate problems with such an argument. Firstly, stable extremal black holes do not seem problematic once one accounts for charge quantization. This ensures that below any given mass $M$ there are a finite number of allowed extremal black hole solutions, $\sim M/e$ where $e$ is the charge quantum. The density of states remains finite therefore, in contrast to the case of black hole remnants. (I will return, however, to the issue of entropy bounds in section~\ref{sec:ultimate}.) Secondly, even in the absence of fundamental charged particles, corrections to pure Einstein-Maxwell theory can lead to deviations from the extremal bound $M \geq Q$ which would allow large black holes to decay into smaller ones. In particular, in ~\cite{Kats:2006xp} it was shown that adding higher-derivative terms to an effective Einstein-Maxwell Lagrangian with signs consistent with UV completion implies that the extremal bound weakens for small black holes, so that $M < Q$ is allowed. These small black holes are then sufficient to allow larger black holes to decay, without invoking fundamental (point-like) particles.

Despite the above counterarguments, one may simply choose to assume as a conjecture that all black holes must be able to decay into sub-Planckian particles (i.e. non-BHs). Then the existence of at least one particle with $m < q$ is required. Analogous conditions have in fact been derived in the context of 3D gravity theories dual to 2D CFTs~\cite{Benjamin:2016fhe,Montero:2016tif}. However, this does not place any practical constraint on EFTs, as it could be satisfied by a particle with mass above the EFT cutoff (even as high as the Planck mass), which would have negligible effects on the low-energy dynamics.~\footnote{It has been argued in~\cite{Ooguri:2016pdq, Freivogel:2016qwc} that the existence of branes or domain walls satisfying WGC-like bounds, even if they are very heavy, imply the instability of non-supersymmetric AdS vacua. Since vacuum stability is a UV-sensitive property, this is consistent with the statement that the minimal WGC doesn't constrain EFT. However, exact stability is crucial for the existence of CFT duals of AdS spacetimes~\cite{Horowitz:2007pr,Harlow:2010az}.} To constrain the content of EFTs, one must adopt a more constraining form of the WGC. Two possibilities were discussed in~\cite{WGC}. One proposal is the so-called ``strong form" of the WGC, which requires that $m < q$ for the lightest charged particle in the spectrum. Another is what I will call the ``unit-charge" WGC, which requires that there exist a particle with the minimal quantum of charge, denoted $\eg$, with mass $m < \eg \mpl$. This would for example imply that any EFT with a $U(1)$ gauge field but no charged particles cannot be valid beyond the scale $\eg \mpl$.

Another more recent proposal, dubbed the ``Sublattice Weak Gravity Conjecture" (sLWGC), was made in ~\cite{Heidenreich:2016aqi,Montero:2016tif}. This proposes that, given a theory with some lattice of possible $U(1)$ charge assignments for states (i.e, a $N$-dimensional lattice if there are $N$ $U(1)$ fields), there is some sublattice with the property that there exists a charged particle for every site on the sublattice, with the norm of the charge vector being greater than the mass. (Earlier it had been proposed~\cite{ Heidenreich:2015nta} that this was true for the entire charge lattice (the ``Lattice WGC" or LWGC), but this was shown to be violated by Kaluza-Klein reduction in certain constructions~\cite{Heidenreich:2016aqi}.) For the case of a single $U(1)$ field, the sLWGC amounts to the statement that there exists some integer $k$ such that for all integers $n$ there exist particles with charge $q_n = n k \eg$ with masses $m_n < q_n$. This infinite tower of particles implies an effective cutoff for the 4D gauge EFT at $\Lambda \sim k \eg \mpl$. However, without any bound on the integer $k$ (defining the size of the sublattice), the sLWGC does not give any bottom-up constraint on EFT. In particular, the examples I will discuss in section~\ref{sec:higgs} will produce low-energy theories which satisfy the sLWGC, but with parametrically large $k$, so that the EFT cutoff is much higher than the na\"{i}ve estimate.       
  
The various forms of the electric WGC are summarized in Table~\ref{tab:wgc} (along with the magnetic WGC). 

Note that none of the electric forms of the WGC discussed here rule out the extranatural inflation model, as they are all satisfied by the light (potentially massless) charged particle in that model, contrary to some claims in the literature~\cite{Rudelius:2014wla,Rudelius:2015xta,Brown:2015iha,Brown:2015lia,Junghans:2015hba,Heidenreich:2015wga}. I review this issue in Appendix~\ref{sec:app}.

\begin{table}
\begin{spacing}{1.5}
\noindent\makebox[\textwidth]{
\begin{tabular}{|>{\raggedright}p{3cm}|>{\raggedright}p{4.8cm}|>{\raggedright}p{5cm}|p{2.3cm}|p{2.1cm}|}

\hline
\bf{Conjecture} & \bf{Statement} & \bf{Motivated by} & \bf{Constrains EFT?} & \bf{Survives Higgsing?}
\\ \hline
Minimal electric WGC & There exists some particle with $q/m > 1$ & Requiring black holes to decay into particles & \begin{center}N\end{center} & \begin{center}Y\end{center}
\\ \hline
Sublattice WGC~\cite{Heidenreich:2016aqi} & For some $k \in \mathbb{Z}$ and all $n \in \mathbb{Z}$, there exist particles of charge $q = n k \eg$ with $q/m > 1$ & Examples in Kaluza-Klein theories and string theory & \begin{center}N\end{center} & \begin{center}Y\end{center}
\\ \hline
Strong electric WGC & The \emph{lightest} charged particle has  $q/m > 1$ & Examples in string theory & \begin{center}Y\end{center} & \begin{center}N\end{center}
\\ \hline
Unit-charge electric WGC & A particle with unit charge has $q/m > 1$  & Examples in string theory & \begin{center}Y\end{center} & \begin{center}N\end{center}
\\ \hline
Magnetic WGC & There is a cutoff of the $U(1)$ theory at a scale $\Lambda \lesssim \eg \mpl$ & Requiring that there exists a monopole which is not a black hole & \begin{center}Y\end{center} & \begin{center}N\end{center}
\\ \hline
\end{tabular}
}
\end{spacing}

\caption{
Summary of the various forms of the WGC discussed in this section, including their definition, motivation, whether or not they place constraints on low-energy EFT models, and whether or not they remain satisfied when a UV gauge group is Higgsed to a smaller one in the IR (as discussed in section~\ref{sec:higgs}). Note that none of the WGC variants which constrain EFTs are robust against Higgsing.}

\label{tab:wgc}

\end{table}

\subsection{``Magnetic'' form}

\label{sec:magnetic}

A stronger constraint on EFT than any of the above conjectures is provided by the magnetic WGC, which states that a $U(1)$ gauge theory with charge quantum $\eg$ must have a cutoff $\Lambda \lesssim \eg \mpl$. The name refers to arguments based on magnetic charge which motivate it, which I now review.

Because $U(1)$ gauge fields should be compact in quantum gravity~\cite{Banks:2010zn}, one can write down magnetically charged black hole solutions, with the magnetic charge obeying the Dirac condition $Q_m = \frac{2 \pi n}{\eg}$ for $n \in \mathbb{Z}$. In particular the minimal (extremal) magnetically charged black hole has $M = Q_m = 2\pi/\eg$, which for a perturbative electric theory is parametrically larger than the Planck scale and therefore na\"{i}vely in the realm of validity of EFT. 

The existence of such a black hole solution may suggest that there should also exist a magnetic monopole which is not a black hole. One could argue this based on the issue of stable black holes~\cite{WGC}, as in the electric case, though this is subject to the same counterarguments regarding the finiteness of the number of stable states. However, in~\cite{delaFuente:2014aca} a qualitatively different argument was suggested. The minimal, extremal magnetically charged black hole has zero temperature and a parametrically large entropy $S \sim 1/\eg^2$, so it actually corresponds to a large number of degenerate quantum states. It seems unusual if magnetic charge can \emph{only} ever appear in such highly entropic configurations. A more typical scenario is for charges to originate from some non-gravitational physics as ``fundamental" (low-entropy) states. If we adopt this as a principle, then we require that there must exist some magnetic monopole that is \emph{not} a black hole.  

With this assumption one obtains the magnetic WGC following~\cite{delaFuente:2014aca}. Magnetic monopoles are necessarily extended objects; i.e. at scales shorter than some length $L$ they are not described by pure $U(1)$ gauge theory. One may therefore expect that this scale represents the cutoff of the $U(1)$ gauge EFT, $\Lambda \sim 1/L$. We can estimate the mass of the monopole from the energy of the magnetic field outside its core, giving $M_\text{monopole} \sim 1/\left(\eg^2 L\right)$. Requiring that the monopole be larger than the Schwarzschild radius associated with this mass, $L \gtrsim \frac{1}{\eg^2 L \mpl^2}$, one obtains the bound $1/L \lesssim \eg \mpl$. The assumption that $\Lambda \sim 1/L$ then gives the magnetic WGC $\Lambda \lesssim \eg \mpl$ However, in section~\ref{sec:higgs} I will show an explicit example in which the above reasoning fails, even if one accepts the premise that there must exist a non-BH monopole. 

Note that due to the inherent ambiguity in the definition of a ``cutoff scale," the magnetic WGC can only be considered as a rough, parametric bound ($\lesssim$ rather than $\leq$), so I have not and will not keep track of $O(1)$ factors when discussing it. This is in contrast to the electric WGC for which the exactness of the bound $Q \leq M$ for black holes leads to a sharp bound on $q/m$ for charged particles.

\subsection{The generalized (multi-field) WGC}
\label{sec:multi}

The above arguments can be extended to theories with more than one $U(1)$ gauge field, giving rise to ``generalized" WGCs (which I will refer to as ``multi-field" WGCs). This was done for the electric WGC in~\cite{Cheung:2014vva}, which considered the necessary conditions for black holes of arbitrary charges under $N$ different $U(1)$'s to decay completely into charged particles. If each particle is associated with a vector $\frac{1}{m}\left(q_1, q_2, ..., q_N  \right)$ where $q_i$ is its charge under the $i$th of the $N$ $U(1)$'s, then the requirement is that the convex hull of all such vectors encloses the unit $N$-sphere~\cite{Cheung:2014vva}. For the simple case of $N$ identical $U(1)$'s, each associated with one charged particle of mass $m$ and charge $q$, this gives a bound of $m \leq q/\sqrt{N}$.   

A similar generalization of the magnetic WGC argument is to demand that a monopole with charge under all $N$ $U(1)$'s is not a black hole. The self-energy of such a monopole is about $\Lambda\left( \frac{1}{g_1^2} + \frac{1}{g_2^2} + ... + \frac{1}{g_N^2}\right)$, so that the requirement that the monopole is larger than its Schwarzschild radius becomes $\Lambda \lesssim \left( \frac{1}{g_1^2} + \frac{1}{g_2^2} + ... + \frac{1}{g_N^2}\right)^{-1/2} \mpl$. If all the gauge couplings are identical, this becomes 
\begin{align}
\label{eq:multi}
\Lambda \lesssim \frac{g}{\sqrt{N}} \mpl.
\end{align} 
Due to the aforementioned $O(1)$ ambiguity of the magnetic WGC bound, this claim is not very meaningful if $N$ is of order a few, but does have import if we consider parametrically large $N$. 

The same result is obtained through a top-down approach in Ref.~\cite{Harlow:2015lma} for emergent gauge fields arising from a $\mathbb{CP}^{N-1}$ model. In that model a gauge field with coupling $g$ emerges from a sector with $\sim 1/g^2$ degrees of freedom, so multiple gauge fields require at least $N_\text{tot} \sim \left( \frac{1}{g_1^2} + \frac{1}{g_2^2} + ... + \frac{1}{g_N^2}\right)$ degrees of freedom. This renormalizes the Planck scale relative to the cutoff as $\mpl \gtrsim \Lambda N_\text{tot}^{1/2}$, giving the above result.

It is clear that arguments against the validity of the single-field WGCs within effective field theory would also rule out these multi-field generalizations. However, in section~\ref{sec:ultimate} I will show that taking the multi-field WGC to hold at some UV scale ensures that the low-energy EFTs satisfy well-motivated entropy bounds. 

\section{Higgsing to Violate the WGC}
\label{sec:higgs}

I now ask the question: If a theory satisfies some form of the Weak Gravity Conjecture, will any low-energy EFT limit of it also satisfy the same WGC?  If not, then regardless of whether this WGC is true for any UV quantum gravity theory, we cannot invoke it to veto EFT models. Instead however we may ask what would be required to complete a given WGC-violating EFT into a WGC-satisfying theory.

Consider a model with two $U(1)$ gauge fields $A$ and $B$. For simplicity take these two $U(1)$'s to have the same charge quantum, $\eg$. The magnetic WGC then implies a field theory cutoff at or below the scale $\Lambda \sim \eg \mpl$, but we are free to consider an EFT below this scale. We can satisfy the electric WGC by having two particles of charge $(1, 0)$ and $(0, 1)$ under $(A, B)$, with masses less than $\Lambda$. (The multi-field WGC would imply additional factors of $\sqrt{2}$ in these bounds, but in this section this $O(1)$ factor will be irrelevant.) Similarly the sLWGC (in fact even the original LWGC) can be satisfied if there are towers of particles above the scale $\Lambda$, with charges $(n, m)$ and mass less than $\sqrt{n^2+m^2} \Lambda$ for all $n, m \in \mathbb{Z}$.

\begin{figure}
\begin{center}
  \includegraphics[width=15cm]{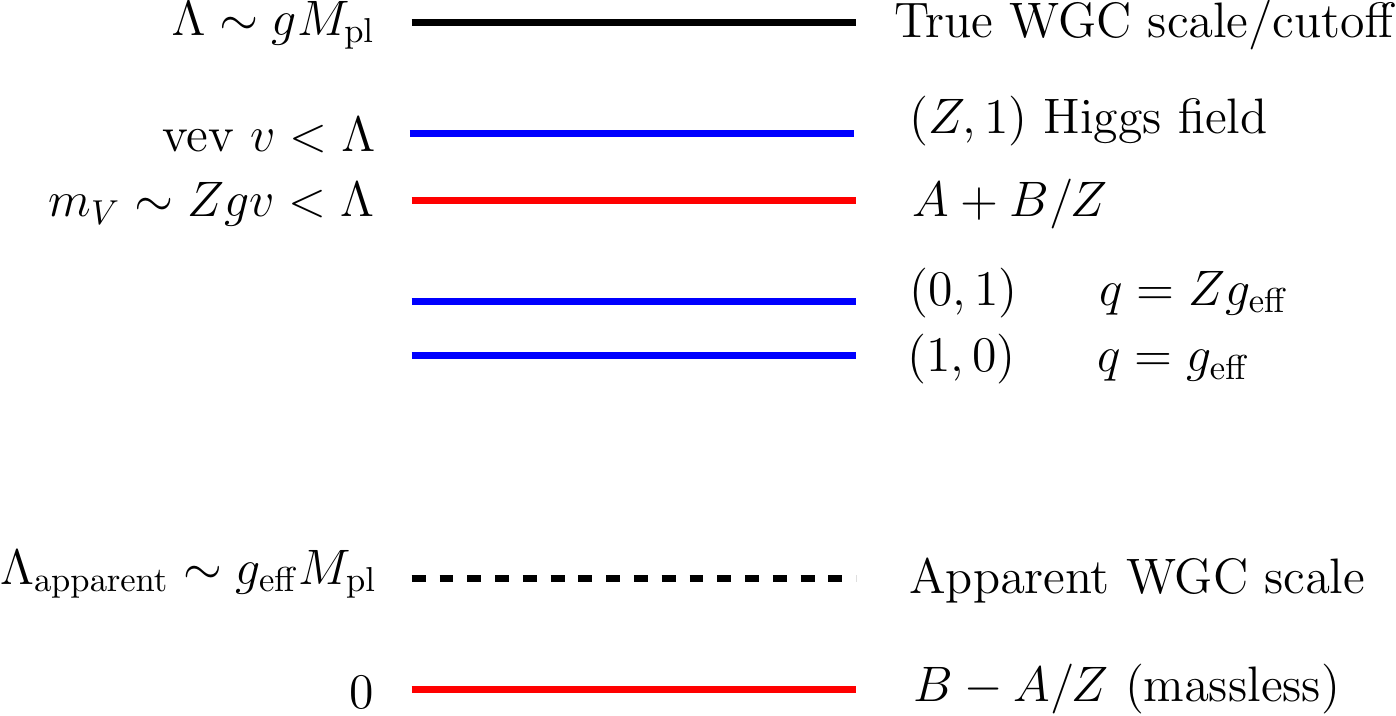}
 \end{center}
 \caption{Spectrum of charged particles (blue) and gauge fields (red) for the model of section~\ref{sec:higgs}, along with the mass scales implied by the WGC (black). This spectrum satisfies the various WGCs for the $A$ and $B$ fields when the scalar of charge $(Z,1)$ does not have a vev. However, when the $(Z,1)$ field acquires a vev, then most forms of the WGC (see Table~\ref{tab:wgc}) are violated for the remaining massless gauge field, which has coupling $\eeff = \eg/Z$.
}
 \label{fig:spec}
 \end{figure}

Now introduce a scalar field with charge $(Z, 1)$, for an integer $Z \gg 1$. Note that if this field is given a mass above those of the $(1, 0)$ and $(0, 1)$ fields, then even the ``strong'' version of the electric WGC is still satisfied. Now consider a phase of this theory where the scalar field acquires a vacuum expectation value $v$, giving a mass $m_V^2 = (Z \eg)^2 v^2$ to the linear combination of fields $V_H \equiv A + B/Z$ by the Higgs mechanism (working everywhere to first order in $1/Z$). The linear combination $V_L \equiv B - A/Z$ remains exactly massless, and at low energies appears as the only observable gauge force. The spectrum of particles in the Higgs phase with their relative masses are shown in Figure~\ref{fig:spec}.
However, charges under the massless $U(1)$ are now quantized in units of $\eeff  \equiv \eg/Z$; e.g. the particle charged as (1, 0) under $(A, B)$ couples with this minimal charge. For parametrically large $Z$, this low-energy gauge coupling can be much smaller than that of the UV theory. Direct application of the magnetic WGC to the low-energy theory would then imply a field theory cutoff at or below $\Lambda_\mathrm{apparent} \sim \eeff \mpl = \eg/Z \mpl$. However it is clear that in the complete theory no actual ``new physics'' is associated with this apparent scale. In particular, the massive gauge field can easily lie well above this scale-- we may take $Z g$ as large as $O(1)$ and $v$ as large as $\Lambda$ in the original theory while maintaining perturbative field theory control, giving $\eeff \equiv \eg/Z \sim \eg^2$ and $m_V \approx \Lambda  \approx \eg \mpl \approx \sqrt{\eeff} \mpl$. Therefore the cutoff of the low-energy $U(1)$ theory can be as high as
\begin{align}
\label{eq:biL}
\Lambda \lesssim \sqrt{\eeff} \mpl
\end{align}
This means that one may have to probe energies well above the apparent magnetic WGC cutoff $\eeff \mpl$ in order to discover the origin of the low-energy theory in terms of two WGC-satisfying gauge fields. Note however that parametric violation of the WGC in the low-energy theory is achieved only if the integer $Z$ in the UV theory is parametrically large.

As for the electric WGC, the weakest form (that there exists some particle with $q > m/\mpl$) is still satisfied in the Higgs theory, as the particle with charges $(0, 1)$ under $(A, B)$ still has charge $\approx \eg$ under the massless gauge field in the Higgs phase and mass $< \eg \mpl$. However, this particle can be heavy enough to be completely invisible in the low-energy theory below $\Lambda_\mathrm{apparent}$. Similarly, if the sLWGC was satisfied in the original theory, then it is also still satisfied in the Higgsed theory, but with a coarser sublattice, e.g. $k = Z$ in the notation of section~\ref{sec:electric}. Again, the particles responsible for satisfying this form of the WGC can all lie well above the scale $\Lambda_\mathrm{apparent}$. In contrast, the forms of the electric WGC which gave meaningful constraints for EFT models, namely the ``strong WGC'' and the ``unit charge WGC'', can both be violated in the low-energy theory, as one can verify for the spectrum of figure~\ref{fig:spec}.

The theory in the un-Higgsed phase satisfied the WGC in all forms, and thus by construction avoided any of the potential ``paradoxes'' for black hole physics. We can ask the same questions about black holes and magnetic monopoles within the WGC-violating EFT after Higgsing. We will find that the paradoxes are all resolved, with the theory essentially exploiting the weaknesses of the arguments discussed in section~\ref{sec:arguments}.

First we can consider the decay of extremal charged black holes, which motivated the electric forms of the WGC. Even in the Higgs phase, black holes with extremal charge under the massless gauge field $B - A/Z$ are guaranteed to be able to decay, due to the existence of the particle with $A, B$ charge $(0, 1)$ with mass $< \eg \mpl$. Such a decay causes black holes to lose $Z$ units of the charge quantum $\geff = \eg/Z$. However, the particle responsible for this decay is much heavier than the na\"{i}ve WGC scale of the low-energy theory, $\geff \mpl$, realizing the ``loophole'' in the minimal electric WGC that prevents it from constraining EFTs. The decay through this particle only allows charges to be lost modulo $g$, however since $g \lesssim 1$ any controlled black hole solution (with $M > Q > 1$ in Planck units) will be able to decay.

\begin{figure}
\begin{center}
  \includegraphics[width=14cm]{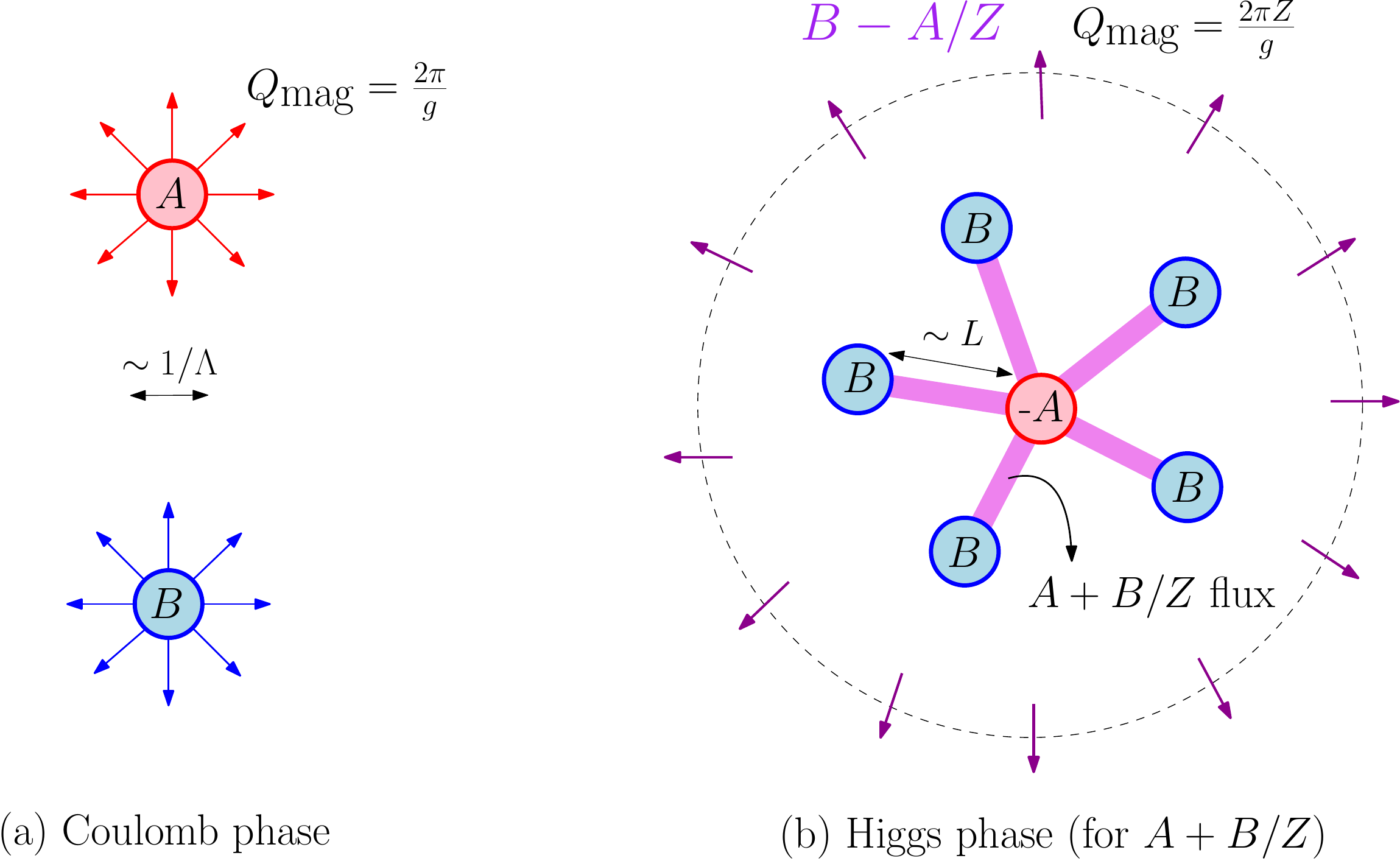}
 \end{center}
 \caption{Magnetic monopoles of the $A, B$ gauge fields in the two phases of the model. Left: In the Coulomb phase, the magnetic WGC requires that magnetic monopoles of both the $A$ and $B$ fields exist, with magnetic charge $2\pi/\eg$ and size of order the cutoff length $\Lambda^{-1}$, where $\Lambda \lesssim \eg \mpl$. Right: In the Higgs phase where a scalar of charge $(Z, 1)$ gets a vev, magnetic flux of the massive gauge boson $A + B/Z$ is confined, and only net $B - A/Z$ flux can escape to infinity. A monopole of $B - A/Z$ charge can be formed by joining $Z$ monopoles of $B$ and one anti-monopole of $A$, with the $A + B/Z$ field confined to flux tubes. 
}
 \label{fig:mp}
 \end{figure}

One can also ask about the status of magnetic monopoles and magnetically charged black holes in the Higgsed theory. Suppose that in the Coulomb phase the theory contains magnetic monopoles of both $A$ and $B$ (Figure~\ref{fig:mp}a), each with charge $2\pi/\eg$, size $\sim \Lambda^{-1}$, and mass $\sim \Lambda/\eg^2$, thereby satisfying the criteria of the magnetic WGC. Higgsing a component of the gauge group then implies confinement of the corresponding magnetic fields (the Meissner effect), such that the magnetic field of the massive gauge boson $A + B/Z$ is confined to flux tubes, while only flux of the massless field $B - A/Z$ can escape to infinity. A long flux tube is a Nielsen-Olesen string~\cite{Nielsen:1973cs} (the analog of Abrikosov vortices in nonrelativistic superconductors) with radius $\sim 1/m_V$ and tension $T \sim v^2$.~\footnote{For definiteness of discussion, I will assume that the mass of the Higgs field is greater than that of the massive gauge boson $m_V \sim Z g v$ (implying that the symmetry-breaking vacuum acts as a type-II superconductor).} Magnetic charges of the massless gauge field are quantized in units of $2\pi/\geff = 2\pi Z/\eg$.  

For example, one could consider a solution with $Z$ monopoles of $B$ and one anti-monopole of $A$, connected by flux tubes (Figure~\ref{fig:mp}b). Since the net magnetic charge under the massive field $A + B/Z$ is zero, such a configuration has finite energy. The size of the system is determined by balancing the energy in the flux tubes $Z v^2 L$ with that of the magnetostatic repulsion of the monopoles $\sim Z^2\left(\frac{2\pi}{\eg}\right)^2 \frac{1}{L}$; in the absence of gravity equilibrium is achieved for $L \sim \frac{\sqrt{Z}}{g v}$. This also determines the mass of the system and thus its Schwarzschild radius, giving $r_S \sim \frac{Z^{3/2} v}{\eg \mpl^2}$. If $r_S/L > 1$ the system in fact a black hole. From the above we have $r_S/L = Z v^2/\mpl^2$. Imposing $v < \Lambda < \eg \mpl$, we have $r_S/L < Z g^2$, which is less than unity for a perturbative theory. Therefore in the regime of EFT control this system is never a black hole; i.e. there exists a magnetic monopole of the unbroken gauge field $B - A/Z$ which is not a black hole, despite the fact that the magnetic WGC is violated for this gauge field!

This example illustrates a specific loophole in the argument presented for the magnetic WGC in the previous section. To obtain the magnetic WGC we demanded that there exist a magnetic monopole which is not a black hole. Since the substructure of a monopole must involve elements beyond the $U(1)$ gauge theory, we interpreted the size $L$ of such a monopole as a cutoff length scale for effective field theory, assuming $\Lambda \lesssim 1/L$. This assumption is true in familiar examples such as the `t Hooft-Polyakov monopole, which is a configuration of some heavy gauge and scalar fields with size controlled by the respective masses. However it is \emph{not} true of the system just discussed, for which the substructure consisted of extended objects (flux tubes) and long range forces. Instead the monopole scale $1/L \sim \frac{g v}{\sqrt{Z}}$ is emergent and is in fact much lower than the scale $m_V = Z g v$ at which new states beyond the low-energy $U(1)$ theory appears. Thus the existence of a non-black-hole monopole does not in fact signal a breakdown of the $U(1)$ gauge theory at a specific scale.

\subsection{Implications for model-building}
\label{sec:model}

The above example indicates that a gauge theory with charge quantum $\eeff$ need not have an EFT cutoff at a scale $\eeff \mpl$ as suggested by the WGC, but instead could be completed into a WGC-satisfying model at scales as high as $\sqrt{\eeff} \mpl$. This possibility of a higher cutoff makes the extranatural inflation model viable. If the above two-field model is realized in the 5D theory, with the inflaton identified as the fifth component of the unbroken gauge field, then the inflaton period is given by $\feff \sim \left( \eeff R \right)^{-1} = \left(\frac{g}{Z} R \right)^{-1}$; imposing that the cutoff scale $\eg \mpl$ is above the compactification scale $1/R$ now gives the bound $f \lesssim Z \mpl$, so that large-field inflation can be achieved. 

This in fact exactly parallels the model proposed in~\cite{delaFuente:2014aca}, which also features two 5D gauge fields $A$ and $B$ and a particle of charge $(Z, 1)$.\footnote{In the notation of~\cite{delaFuente:2014aca}, the large charge we denote as $Z$ appears as $N$, which in this work refers to the number of $U(1)$ gauge fields.} In that model, this particle did not Higgs the $A + B/Z$ gauge field, but instead gave a large mass for the axion in the $A + B/Z$ direction through quantum corrections. As in the model discussed here, this left the $B - A/Z$ direction with a small effective coupling, or large effective axion period (realizing the ``alignment" scenario of~\cite{kimNillesPeloso}). The exact same constraints apply for both models: $1/R \lesssim \eg \mpl$ to satisfy the WGC, and $Z \eg \lesssim 1$ for perturbativity.

We can also express the constraints in these models in terms of the inflationary observables. The number of $e$-foldings that inflation lasts goes as $\nefolds \sim \feff/\mpl$, while the Hubble scale during inflation is $H \sim \sqrt{V}/\mpl \sim 1/\left(\mpl R^2 \right)$. When saturating the constraints, one has $\feff/\mpl \sim Z \sim 1/\eg \sim \mpl R$, or in terms of the observables 
\begin{align}
\label{eq:H}
H \lesssim \frac{\mpl}{\nefolds^2}; 
\end{align}
i.e. a long period of inflation comes at the expense of a low Hubble scale. The cosmological data require $H/\mpl \sim 10^{-4}$ and $\nefolds > 60$. One can see that this is close to saturating the bound of eq.~\ref{eq:H} (though of course this expression lacks $O(1)$ factors). In physical terms this implies that one or more of the constraints for EFT control will be close to being saturated, i.e. that the EFT cutoff $\Lambda$ (which could be either the WGC scale $\eg \mpl$ or the scale at which the 5D gauge theory becomes strongly coupled, $\sim \frac{1}{Z \eg} \frac{1}{R}$) is not far above the compactification scale $1/R$. 

\begin{figure}
\begin{center}
  \includegraphics[width=8cm]{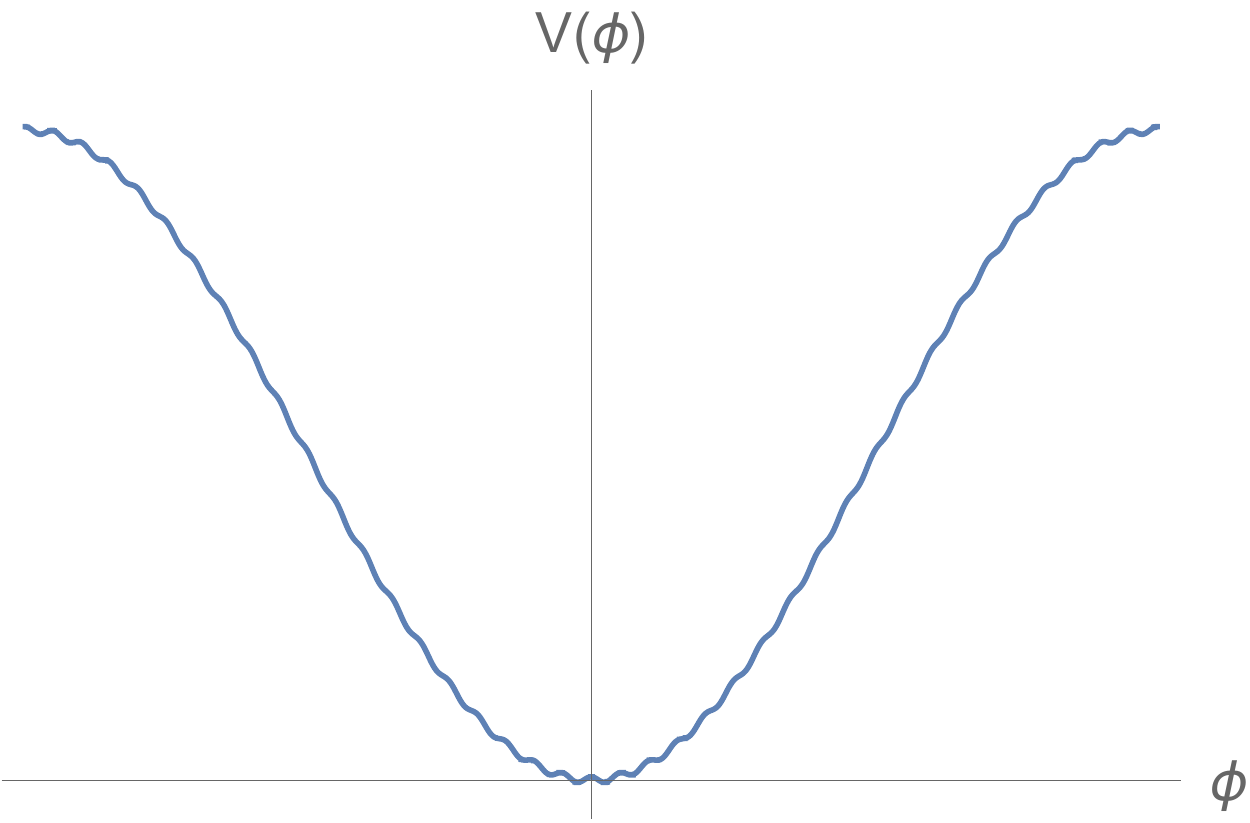}
 \end{center}
 \caption{A sketch of a possible inflaton potential generated within the bi-axion models discussed here and in~\cite{delaFuente:2014aca}, for an EFT cutoff that is only slightly above the compactification scale. In this case the potential may receive a suppressed `higher harmonic" contribution with a period $1/Z$ times the total field range, leading to observable oscillations in the primordial power spectrum. (The amplitude of this contribution is exaggerated in this figure; current data constrains it to be smaller than pictured.)   
}
 \label{fig:osc}
 \end{figure}

A low EFT cutoff has important quantitative implications for the model, as the cutoff scale can generically be associated with the masses of new particles, which may carry gauge charge and thus contribute to the axion potentials. The strength of the extranatural inflation model lies in its relative insensitivity to UV physics at or above the cutoff: because the axions are emergent fields in 4D, corresponding to nonlocal Wilson loop operators in the 5D theory, any unknown physics at scales smaller than the compactification scale gives suppressed contributions to the axion potential. For example, charged particles with 5D mass $m_5$ contribute to the axion potentials with a factor of $e^{-2 \pi R m_5}$, which can be thought of as the Yukawa suppression for a massive particle to propagate around the $S^1$~\cite{ArkaniHamed:2007gg}. For $m_5$ not too far above $1/R$ however, this correction to the potential, though suppressed, may still be relevant for phenomenology. In particular, a particle of charge $(0, 1)$ under $(A, B)$ would give a contribution to the inflaton potential with period $\frac{1}{2\pi R \eg } = \feff/Z$, i.e. much shorter than the desired slow-roll potential. The result would be a slow-roll potential modulated by small-amplitude, high-frequency oscillations as sketched in Figure~\ref{fig:osc}. Observationally this implies an oscillating component in the scalar power spectrum, as discussed in ~\cite{Wang:2002hf, Pahud:2008ae,Flauger:2009ab,2011JCAP...01..026K,Easther:2013kla,Flauger:2014ana}; in~\cite{delaFuente:2014aca} it was discussed how these bi-axion models may be constrained or probed by searches for such oscillations in the current data (most recently~\cite{Flauger:2014ana}).

The above model with two $U(1)$'s was conceived as a minimal way to start from a UV theory that satisfies the WGC and produce a low-energy theory that violates it, as is necessary for extranatural inflation to be realized. In this model however the EFT cutoff was still bounded by eq.~\ref{eq:biL}, implying that large-field inflation is possible but constrained, i.e. by eq.~\ref{eq:H}. However, by considering more fields one can realize models with even weaker constraints on the EFT cutoff and therefore on phenomenology. For example, one can essentially iterate the above Higgsing pattern with three gauge fields $A, B, C$ and two Higgs fields with charges $(Z, 1, 0)$ and $(0, Z, 1)$. (As before, this is exactly paralleled in the ``tri-axion" models of ~\cite{yangBai, delaFuente:2014aca}, without Higgsing.) Now the effective gauge coupling of the massless $U(1)$ is $\eeff = \eg/Z^2$, in terms of which the cutoff bound is expressed as $\Lambda \lesssim \left(\eeff\right)^{1/3} \mpl$ rather than eq.~\ref{eq:biL}. This allows even larger value for the extranatural inflation field range, $\feff \sim \left( \eeff R \right)^{-1}$, and eliminates any tension between EFT control and fitting the cosmological data~\cite{delaFuente:2014aca}. 

In these models, avoiding the constraints and signals na\"{i}vely suggested by the WGC clearly came at the cost of some model-building complexity. The original extranatural model with one gauge field could not realize any period of inflation consistent with the WGC. The bi-axion models are the most minimal models which could produce inflation while obeying the WGC (in the full theory); while these can fit the current data, they could potentially be ruled out by further measurements. However, one could still ``model-build around" any falsification by proceeding to more non-minimal models with additional gauge fields and specific patterns of charge assignments, still satisfying the WGC. Strictly speaking therefore, even if one takes the WGC to hold at some UV scale, one does not obtain an inevitable prediction (of the form of Eq.~\ref{eq:H} or the oscillations of Fig.~\ref{fig:osc}) from the extranatural inflation framework. However, if one invokes minimality (of field content and/or charge assignments) as a model-building criterion, then imposing the WGC implies that the bi-axion models are favored and their predictions should be considered as important benchmarks in testing extranatural inflation. The WGC can therefore still have some relevance in guiding model-building despite not being a true constraint at the EFT level. In the concluding section~\ref{sec:conc} I will explore this theme further.

\section{A Proposed Bound on Gauge EFTs}
\label{sec:ultimate}

Repeated Higgsing of gauge fields after the pattern of the previous section allows for exponentially small gauge couplings in low-energy EFT, even if the UV theory satisfies the Weak Gravity Conjecture. This is in direct analogy to axion models which generate exponentially large decay constants through alignment of multiple fields~\cite{axion1,Higaki:2014pja,Choi:2015fiu,Kaplan:2015fuy}. To be concrete, consider $N$ Abelian gauge groups $U(1) \times U(1) \times ... \times U(1)$ all with coupling $\eg$, and $N-1$ Higgs fields with charges $(Z, 1 , 0, ..., 0)$, $(0, Z, 1, 0, ..., 0)$, ... $(0, ..., Z, 1)$. (Taking these couplings and charges to be equal maximizes the violation of the WGC at low energy.) When all of these Higgs fields acquire vevs, the remaining massless $U(1)$ has charge quantum $\eeff = \eg/Z^{N-1}$. After applying the perturbativity bound $Z g \lesssim 1$ we have the bound $\eeff \gtrsim \eg^N$. Therefore exponential violation of the WGC can be generated even if the integers $Z$ and $N$ are order a few. However, there is still a limit on the coupling that can be achieved, as when the multi-field WGC is taken into account then large $N$ also lowers the cutoff as $\Lambda \lesssim \frac{\eg}{\sqrt{N}} \mpl$. In terms of the low-energy coupling, the cutoff bound is $\Lambda \lesssim \frac{\eeff^{1/N}}{\sqrt{N}} \mpl$. $N$ can be chosen to maximize this upper bound, which gives 

\begin{align}
\label{eq:ult}
\Lambda \lesssim \left(\log \frac{1}{\eeff}\right)^{-1/2} \mpl.
\end{align}
That is, if a theory satisfies the multifield WGC in the UV, then repeated Higgsing cannot generate an EFT where the effective gauge coupling $\eeff$ violates this bound. 

Interestingly, a completely separate argument for equation~\ref{eq:ult} can be made based on considerations of entropy bounds , which I will now present. This draws on similar reasoning as the argument against exact global symmetries in quantum gravity: If black holes could carry arbitrary, conserved global charges, then they could store an infinite amount of information, in violation of entropy bounds-- e.g., the holographic entropy bound~\cite{Fischler:1998st,Bousso:1999xy} indicating that the entropy of a volume of area $A$ is given by $S = A/4G$. Gauge symmetries are not ruled out in the same way since they produce long-range forces which affect the structure of black holes; in particular black holes can only carry a finite amount of charge, $Q \leq M$. However, consider the set of charged black hole solutions of radius $R$, which have $M \sim R$ and $Q \leq M$. Suppose that all charged particles have mass well above the BH temperature $T \sim R^{-1}$, so that black holes of different charge should be treated as separate thermodynamic ensembles. If the quantum of charge is $\eeff$, then there are $\sim R/\eeff$ thermodynamically distinct charged black holes in this set, each with a Bekenstein-Hawking entropy of $S_\text{BH} = A/4 \sim R^2$. (We could imagine creating all of these black holes by dropping charges into a large Schwarzschild black hole and letting it evaporate to radius $R$.) So the total number of distinct states in this set is in fact $e^{S}$ where $S = S_\text{BH} + \log\left(R/\eeff\right)$. The first term saturates the usual holographic entropy bound, while the second can be thought of as a correction that is typically well subdominant (e.g. for $\eeff \sim O(1)$, $R \gtrsim \mpl^{-1}$). However, for exponentially small $\eeff$, there is some $R$ for which the second term in fact dominates the entropy, namely if $R^2 \lesssim \log\left(R/\eeff\right) \approx \log\left(1/\eeff\right)$. If we wish to avoid gross violation of the entropy bound, then we must introduce some new physics to this setup at distance scales larger than this, i.e. at some energy scale 
\begin{align*}
\Lambda \lesssim \left(\log \frac{1}{\eeff}\right)^{-1/2} \mpl.
\end{align*}
Note that this ``new physics'' could perhaps be as mild as introducing a charged particle below this scale, which could be thermally Hawking radiated from black holes of this size. Then black holes of different charge should not be considered as distinct thermodynamic ensembles when calculating the Bekenstein-Hawking entropy. Alternately, the Einstein-Maxwell semiclassical description of black holes could break down in some way at this scale.

I propose that eq.~\ref{eq:ult} is the only version of the Weak Gravity Conjecture which can be considered as a constraint on low-energy EFTs, due to the bottom-up argument just presented. It is interesting and nontrivial that this bound can be \emph{saturated but not violated} when Higgsing theories which satisfy the multifield WGC in the UV. Following the swampland philosophy, one may have expected that not all possibilities for field content are consistent within quantum gravity, such that it could be possible to write down a Higgs model that produces a gauge coupling small enough to violate eq.~\ref{eq:ult}. However, we instead find that, given that the theory at some scale satisfies the multifield WGC as expressed by eq.~\ref{eq:multi}, one can consider arbitrary matter fields without being in danger of landing in an ``EFT swampland" by violating eq.~\ref{eq:ult}. 

\section{Conclusions}
\label{sec:conc}

In this work I have considered whether the Weak Gravity Conjecture could be a robust constraint on effective field theories, particularly in the context of EFT models generated by Higgsing theories with multiple gauge fields. I showed that a UV theory which satisfies the WGC could have a low-energy EFT description which badly violates the usual WGC bounds. However, such an EFT always satisfies the much weaker constraint of eq.~\ref{eq:ult}. This latter bound is also motivated by requiring that charged black holes do not grossly violate the holographic entropy bound.    

One may ask whether these Higgsing models actually \emph{disprove} the forms of the WGC which constrain EFT (the last three entries of Table~\ref{tab:wgc}). These conjectures could still hold, even in a theory with a large landscape of EFTs, if the fundamental theory is such that the patterns of Higgsing I discussed are never realized. This could occur for example if scalar fields are never realized with the gauge charges I considered. While this is in principle a possibility, it would require the WGC to be expanded to a much more extensive (i.e, more hypothetical) conjecture, with no additional motivation. Indeed, as I have discussed, the original bottom-up arguments for the WGC based on black hole physics are all evaded in the Higgsing model in well-understood ways. Note however that if we allow arbitrary matter representations to Higgs the gauge fields, then the constraint implied by entropy bounds, eq.~\ref{eq:ult}, is guaranteed if and only if the theory \emph{does} satisfy the multifield WGC at some scale. 

Taken together, these results suggest the following interpretation of the physics surrounding the Weak Gravity Conjecture: the conventional forms of the WGC may always be satisfied at some UV scale, perhaps the scale at which gauge field theory is completed into some quantum gravity theory. Consistent with this one could consider arbitrary additional matter fields which spontaneously break these symmetries as I have discussed, giving emergent violation of the WGC at the low-energy EFT level. However, due to the ``boundary condition" of having the WGC realized in the UV, there never occurs a problematic situation in which entropy bounds are violated; i.e. eq.~\ref{eq:ult} is always satisfied.  

This viewpoint reconciles the arguments for violation of the WGC presented here with the numerous examples of the WGC being satisfied in string theory~\cite{WGC, Hebecker:2015zss} and other models where gauge fields emerge from more fundamental objects~\cite{Heidenreich:2015nta, Harlow:2015lma}. Furthermore, it suggests that parametric violation of the WGC in EFT, while possible, could be considered to be ``non-minimal."  In the models I discussed, achieving significant violation of the WGC at low energies required specific field content and interactions, namely parametrically large integer charges and/or particular textures of charges for multiple fields. These ingredients could appear at very high scales, perhaps close to the field theory cutoff, but were necessary at some point if the UV theory satisfied the WGC. Increasingly strong violations of the WGC required additional fields and more specific charge assignments. One may expect that this specificity makes such models ``unlikely"; for example in the context of theories with a large landscape of low-energy EFTs one might expect that vacua with strong violation of the WGC are the exception rather than the norm (under some suitable measure). Of course, such conclusions are not rigorous as the concept of minimality cannot be uniquely defined-- perhaps some deeper but simpler structure could underlie what appears to be a strange proliferation of fields and charges. Nevertheless, these considerations could be taken as reasonable motivations to guide model-building.   

As stressed in the introduction (sec.~\ref{sec:intro}), the status of the WGC is an issue of practical importance in model-building for inflation etc. The results of this work indicate that the usual WGC constraints cannot be considered as absolute rules for EFT models, though the bound of eq.~\ref{eq:ult} perhaps should be. This inequality is so much weaker than the WGC though so as to be almost irrelevant (though it may still be important for models such as the relaxion~\cite{Graham:2015cka} which may invoke field ranges many many orders of magnitude greater than the Planck scale). However, taking the viewpoint discussed in this section, one could seek to complete WGC-violating models into WGC-satisfying theories as minimally as possible. For example, as discussed at the end of section~\ref{sec:model}, applying these considerations to extranatural inflation focuses one's attention on bi-axion models and their specific predictions for cosmological observables, such as oscillations in the power spectrum. In this sense the WGC, even when imposed only in the UV and not at the low-energy EFT level, \emph{does} guide model-building and phenomenology in a meaningful way.

\section*{Acknowledgments}

I am grateful to Anton de la Fuente, David E. Kaplan, Jared Kaplan, and Raman Sundrum for useful discussions. I especially thank Anton and Raman for early collaboration on aspects of this work; in particular the arguments reviewed in the appendix were originally developed in collaboration following our work in~\cite{delaFuente:2014aca}. This research was supported by the National Science Foundation grants PHY-1315155 and PHY-1214000 and by the Maryland Center for Fundamental Physics.

\appendix
\section{The Electric WGC and Large Field Inflation}
\label{sec:app}

In this appendix I review why the electric forms of the WGC do not constrain large-field axionic inflation models such as extranatural inflation, contrary to the claims or suggestions of~\cite{Rudelius:2014wla,Rudelius:2015xta,Brown:2015iha,Brown:2015lia,Junghans:2015hba,Heidenreich:2015wga}. These arguments, originally pointed out by the authors of~\cite{delaFuente:2014aca} and Matthew Reece~\cite{reece}, have been previously referenced in~\cite{Rudelius:2015xta} and later~\cite{Heidenreich:2015wga,Brown:2015lia}; I review them here so as to present a complete discussion of the implications of the WGC for model-building. 

If the inflaton is realized as an axion, it must have a flat potential over some effective field range $f_\text{eff} \gtrsim \mpl$. The potential of an axion arises from nonperturbative effects such as instantons, generally characterized by some instanton action $S$. For example, the contribution of a charged particle to the potential of the inflaton in extranatural inflation can be calculated in the worldline formalism which involves the action of a charged particle propagating in a loop around the extra dimension. The contribution of an instanton of action $S$ to the axion potential can be written in the general form 
\begin{align}
\label{eq:Vax}
V \sim V_0 \sum_n c_n e^{-n S} \cos \frac{n \phi}{f}
\end{align}
which represents a sum over configurations with $n$ instantons. For extranatural inflation, a charged particle of mass $m$ and charge $q$ gives $S = 2\pi m R$ for the particle propagating around the extra dimension, $f = \left(2 \pi R q\right)^{-1}$, and $c_n \approx 1/n^5$ for $m \ll R^{-1}$~\cite{Hosotani,massiveHosotani}. For more general axions however the $c_n$ may not be readily calculable. If the charged particle satisfies the electric WGC, $q/m > 1$, then the potential parameters satisfy 
\begin{align}
\label{eq:0form}
S < \mpl/f
\end{align}

In~\cite{WGC} it was proposed that the above relation is satisfied for axions in more general contexts, constituting a ``0-form" version of the WGC. The arguments of~\cite{Rudelius:2014wla,Rudelius:2015xta,Brown:2015iha,Brown:2015lia,Junghans:2015hba,Heidenreich:2015wga} against large field axion inflation invoke generalized multi-field versions of this 0-form WGC and show that, even in the case of a multidimensional axion field space, realizing an effective decay constant $f_\text{eff} > \mpl$ for some direction requires the existence of an instanton with $S < 1$. But $S < 1$ in a potential of the form~\ref{eq:Vax} implies that the higher harmonic terms (with $n>1$) are no longer parametrically suppressed. In general this may prevent any theoretical control of the potential, or at least spoil its flatness and prevent inflation. 

However, this last conclusion does not always follow-- in fact, extranatural inflation provides a clear counter-example. Even for a massless charged particle, trivially satisfying the WGC and giving $S = 0$ in eq.~\ref{eq:Vax}, the potential can be explicitly calculated in perturbation theory, and has the higher harmonic terms (coefficients of $\cos \frac{n \phi}{f}$) suppressed by factors of $c_n = 1/n^5$. This is in fact enough to to ensure that the potential is convergent and in fact nearly indistinguishable from a pure cosine in practice. In the models considered here and in the literature~\cite{yangBai, delaFuente:2014aca} the 5D gauge theory is always taken to be perturbative, ensuring that the loop expansion of the potential is valid. Thus there is sufficient theoretical control to guarantee successful inflation, despite the lack of \emph{parametric} suppression of higher harmonics.

In fact it is even possible to achieve an axion potential \emph{with} parametric suppression of higher harmonics consistent with a minimal version of the 0-form WGC, due to another loophole in these arguments. Suppose that more than one instanton contributes to the axion potential, giving contributions of the form of eq.~\ref{eq:Vax} but with distinct actions $S_i$ and decay constants $f_i$. The 0-form analog of the minimal electric WGC is the requirement that there exists \emph{some} instanton $S_\text{WGC}$ with corresponding decay constant $f_\text{WGC}$ satisfying $S_\text{WGC} < \mpl/f_\text{WGC}$ (corresponding in extranatural inflation to having some particle with $m < q$). However, one could have a situation where some other instanton $i$ that does \emph{not} satisfy the WGC has a smaller action, i.e. $S_\text{WGC} \gg S_i > 1$, with $f_i > \mpl$. The WGC-satisfying instanton gives a contribution to the potential with period $f_\text{WGC} \ll \mpl$, but this is exponentially subdominant to the contribution from $S_i$ with super-Planckian period, so that the potential can be flat enough to support inflation. The lack of constraint exactly parallels the fact that the minimal (1-form) electric WGC cannot constrain EFT models at all, due to the possibility that the particle satisfying $m < q$ can still have a mass far above the scales of relevance for the EFT. Constraints on axion inflation can only be obtained by invoking stronger versions of the 0-form WGC analogous to the last three entries of Table~\ref{tab:wgc}, but as discussed here and elsewhere these forms of the WGC do not seem to be theoretically robust.

\bibliography{wgcbib}

\end{document}